# Taylor series of Landauer conductance

Carlos Ramírez*, Mauricio J. Rodríguez, Bryan D. Gomez

*Departamento de Física, Facultad de Ciencias, Universidad Nacional Autónoma de México, Apartado Postal 70542, 04510 Ciudad de México, México*

*Corresponding author. E-mail address: carlos@ciencias.unam.mx

**Abstract**

In this paper, we propose a method to calculate the exact Taylor series of the scattering matrix in general multiterminal tight-binding systems to arbitrary order N, which allows us to find the Taylor expansion of Landauer conductance in mesoscopic systems. The method is based on the recursive scattering matrix method (RSMM) that permits us to find the scattering matrix of a system from the scattering matrices of its subsystems. Following ideas of automatic differentiation, we determine expressions for the sum, product, inverse, and diagonalization of a matrix Taylor expansion, and use them into the RSMM to find Taylor series of scattering matrices. The method is validated by obtaining the transmission function of atomic chains with site defects and graphene nanoconstrictions. Finally, an analysis of convergence radius and error estimations of these Taylor expansions is presented.

## 1. Introduction

In the last decades, advances in the miniaturization of technological devices have reached the mesoscopic scale, where coherence lengths are larger than system size, so interference quantum phenomena must be deeply understood. At zero temperature, electronic conductance $(G)$ in these structures can be obtained from the celebrated Landauer formula [1–3],

$$G = \frac{2e^2}{h} T(E_F), \tag{1}$$

by means of the transmission function $(T)$ evaluated at the Fermi energy $(E_F)$. Conductance is strongly dependent on impurities, defects, orderings, and geometries. For example, a Fibonacci sequence of site impurities in a tight-binding atomic chain has a fractal-like conductance spectrum [4], while a periodic sequence of impurities generates an oscillatory spectrum.

Ordinarily, conductance (1) is calculated using the recursive Green function method. Also, it can be obtained straightforwardly from the scattering matrix (S-matrix),

$$\mathbf{S}(E) = \begin{pmatrix} \mathbf{r}(E) & \mathbf{t}'(E) \\ \mathbf{t}(E) & \mathbf{r}'(E) \end{pmatrix}, \tag{2}$$

as $T(E_F) = \text{Tr}\left[\mathbf{t}^\dagger(E_F)\mathbf{t}(E_F)\right]$. In tight-binding Hamiltonians, S-matrices can be efficiently computed using the recursive scattering matrix method (RSMM) [5,6]. In any case, the conductance is usually solved numerically by evaluating (1) at discrete values of energies. However, if the set of these energies is not dense enough, details of the conductance spectrum may be missed, nonetheless, an overly dense evaluation may be computationally expensive without revealing new information.

In general, it would be convenient to obtain $T(E)$ analytically. Remarkably, this has been achieved in some systems, such as atomic chains with impurities [4,7,8], atomic chains with Fano defects [9], atomic chains with attached Aharonov–Bohm loops [10–13], armchair-type carbon nanotubes with site defects [14], armchair graphene nanoconstrictions (GNC) [15], zigzag phosphorene nanoribbons with site defects [16], molecular wires [17,18], graphene nanobubbles [19], graphene heterojunctions [20], graphene nanoribbons with defects [21], and parallel armchair nanotubes [22]. Transparent states have also been analytically demonstrated in atomic chains with impurities following Fibonacci orderings [4], atomic chains with Fano defects [9], and disordered nanotapes with Fano defects [8]. Unfortunately, analytical expressions are not available in general complex structures.

Alternatively, a Taylor series may provide further information about the behavior of conductance in the neighborhood of each energy. By means of perturbation theory, first-order Taylor expansions of Landauer conductance have been useful to analyze Büttiker probes [23], transition voltage spectroscopy [24], inelastic electron tunneling spectra [25], electromagnetic induced transparency and reflection [13], and quantum pumping [26]. These studies could be improved with higher-order Taylor expansions.

In this paper, we explain how to find the order $N$ Taylor series of Landauer conductance in arbitrary tight-binding nanostructures. In section 2, RSMM is briefly described and the method is extended to calculate Taylor expansion of S-matrices with general semi-infinite leads. In section 3, the method is validated in atomic chains with site defects and graphene nanoconstrictions, and the convergence of these Taylor expansions is discussed.

## 2. Recursive scattering matrix method (RSMM)

The RSMM allows us to calculate the S-matrix of a tight-binding system in terms of the S-matrices of its subsystems [5,6]. Examples of these systems are shown in Fig. 1, and they consist of a dispersion region attached to semi-infinite atomic chains (blue dot-dashed lines) with null site energies and hopping integrals $t_c > |E/2|$, where $E$ is the energy.

For a system A, the projection coefficient of the wavefunction into the $m$-th site of the $n$-th coupled chain is

$$a_{n,m} = A_n^{(+)} e^{-i\kappa m} + A_n^{(-)} e^{i\kappa m}, \tag{3}$$

where $A_n^{(+)}$ and $A_n^{(-)}$ are coefficients of incoming and outgoing waves in the $n$-th coupled chain, respectively, and $E = 2t_C \cos(\kappa)$. These coefficients are related by,

$$\mathbf{A}^{(-)} = \mathbf{S}^A \mathbf{A}^{(+)} = \begin{pmatrix} \mathbf{S}_{11}^A & \mathbf{S}_{12}^A \\ \mathbf{S}_{21}^A & \mathbf{S}_{22}^A \end{pmatrix} \mathbf{A}^{(+)}, \tag{4}$$

where $\mathbf{A}^{(+)} \left( \mathbf{A}^{(-)} \right)$ is the vector of coefficients of incoming (outgoing) waves in the coupled chains, and $\mathbf{S}^A$ is the S-matrix of the A system. Analogous equations can be stated for a system B by changing $A$ for $B$ in the previous expressions.

As demonstrated by Ramírez and Medina-Amayo [5], by making $A_n^{(\pm)} = B_n^{(\mp)}$ for $n = 1, \cdots, M$, systems A and B are glued together producing system C, as exemplified in Fig. 1(c). In particular, the S-matrix of system C can be obtained from the S-matrices of systems A and B,

$$\mathbf{S}^C = \begin{pmatrix} \mathbf{S}_{22}^A + \mathbf{S}_{21}^A \left( \mathbf{I} - \mathbf{S}_{11}^B \mathbf{S}_{11}^A \right)^{-1} \mathbf{S}_{11}^B \mathbf{S}_{12}^A & \mathbf{S}_{21}^A \left( \mathbf{I} - \mathbf{S}_{11}^B \mathbf{S}_{11}^A \right)^{-1} \mathbf{S}_{12}^B \\ \mathbf{S}_{21}^B \left( \mathbf{I} - \mathbf{S}_{11}^A \mathbf{S}_{11}^B \right)^{-1} \mathbf{S}_{12}^A & \mathbf{S}_{22}^B + \mathbf{S}_{21}^B \left( \mathbf{I} - \mathbf{S}_{11}^A \mathbf{S}_{11}^B \right)^{-1} \mathbf{S}_{11}^A \mathbf{S}_{12}^B \end{pmatrix}, \tag{5}$$

with $\mathbf{S}_{ij}^{A/B}$ submatrices of $\mathbf{S}^{A/B}$ as shown in Eq. (4), where the S-matrix was divided into sections such as $\mathbf{S}_{11}^{A/B}$ are square matrices of dimension $M \times M$. The equation (5) allows us to recursively find the S-matrix of arbitrary complex tight-binding systems, starting from the S-matrices of the bond and site structures, shown respectively in Figs. 1(d) and (e). The S-matrices of these structures are

$$\mathbf{S}^{bond} = \begin{pmatrix} r & \frac{t_C}{t}\left(e^{i\kappa} + re^{-i\kappa}\right) \\ \frac{t_C}{t}\left(e^{i\kappa} + re^{-i\kappa}\right) & r \end{pmatrix}, \tag{6}$$

with $r = -\left(t^2 - t_C^2\right)/\left(t^2 - t_C^2 e^{-2i\kappa}\right)$, and

$$\left(\mathbf{S}^{site}\right)_{n,m} = \frac{2it_C \sin\kappa}{\varepsilon - E + Pt_C e^{i\kappa}} - \delta_{n,m}. \tag{7}$$

S-matrices are functions of $E$. Let us assume that $\mathbf{S}^A$ and $\mathbf{S}^B$ are given as truncated Taylor series of order $N$,

$$\mathbf{S}^{A/B}(E) = \sum_{n=0}^{N} \mathbf{S}_n^{A/B}\left(E - E_0\right)^n + O\left(E - E_0\right)^{N+1}, \tag{8}$$

where $\mathbf{S}_n^{A/B} \equiv \frac{1}{n!} \partial^n \mathbf{S}^{A/B}/\partial E^n \big|_{E=E_0}$. According to Eq. (5), $\mathbf{S}^C$ can be found by making operations of sum, product, and inverse with submatrices of $\mathbf{S}^A$ and $\mathbf{S}^B$. These operations can be performed using the Taylor series (8), leading to the exact order-$N$ Taylor series of $\mathbf{S}^C$ (see Appendix A). Moreover, S-matrices of site and bond structures can be expressed as Taylor series by using automatic differentiation [27] on each matrix element. Consequently, Eq. (5) allows us to find recursively the S-matrix of tight biding systems as Taylor expansions.

*2.1 The leads*

Within the Landauer formalism, the leads are semi-infinite periodic structures connected to the scattering region. A lead structure is exemplified in Fig. 1(f) and consists of a semi-infinite periodic structure attached to semi-infinite atomic chains (blue dot-dashed lines) with null site energies and hopping integrals $t_c > |E/2|$. Recently, Ramírez proposed a general method to find the S-matrix of an arbitrary lead structure $\left(\mathbf{S}^L\right)$ [6]. In the following, we describe briefly this method.

First, we identify a unit structure that builds the semi-infinite lead when the RSMM is applied infinite times, as the one illustrated in Fig. 1(g). The S-matrix of the unit structure $\left(\mathbf{S}^U\right)$ can also be calculated with the RSMM. $\mathbf{S}^U$ is a $2Q \times 2Q$ matrix, being $Q$ the number of coupled chains to the left, or to the right, of the unit structure, and satisfies

$$\begin{pmatrix} \mathbf{A}^{(-)} \\ \mathbf{B}^{(-)} \end{pmatrix} = \mathbf{S}^U \begin{pmatrix} \mathbf{A}^{(+)} \\ \mathbf{B}^{(+)} \end{pmatrix} = \begin{pmatrix} \mathbf{S}^U_{11} & \mathbf{S}^U_{12} \\ \mathbf{S}^U_{21} & \mathbf{S}^U_{22} \end{pmatrix} \begin{pmatrix} \mathbf{A}^{(+)} \\ \mathbf{B}^{(+)} \end{pmatrix}, \qquad (9)$$

where $\mathbf{A}^{(\pm)}$ and $\mathbf{B}^{(\pm)}$ are vectors of coefficients of right and left coupled chains, respectively. Then, by solving the generalized eigenvalue problem

$$\begin{pmatrix} -\mathbf{S}^U_{11} & \mathbf{I} \\ -\mathbf{S}^U_{21} & 0 \end{pmatrix} \begin{pmatrix} \mathbf{A}^{(+)} \\ \mathbf{A}^{(-)} \end{pmatrix} = \lambda \begin{pmatrix} 0 & \mathbf{S}^U_{12} \\ -\mathbf{I} & \mathbf{S}^U_{22} \end{pmatrix} \begin{pmatrix} \mathbf{A}^{(+)} \\ \mathbf{A}^{(-)} \end{pmatrix}, \qquad (10)$$

we find Bloch states (open channels) if $|\lambda|=1$, left evanescence states if $|\lambda|>1$ and right evanescence states if $|\lambda|<1$. These eigenvalues can be grouped into $Q$ pairs $(\lambda, \lambda')$ that accomplish $\lambda\lambda'=1$. By writing eigenvalues of Bloch states as $\lambda_k = e^{ika}$, where $a$ is the lattice parameter, the band structure of the infinite crystal can be calculated. Equation (10) can be restated as an ordinary eigenvalue problem only if there are non-zero eigenvalues.

This occurs when the chosen unit structure has the minimum $Q$ [6]. In this case, we may alternatively solve

$$\Lambda \begin{pmatrix} \mathbf{A}^{(+)} \\ \mathbf{A}^{(-)} \end{pmatrix} \equiv \begin{pmatrix} 0 & \mathbf{S}^U_{12} \\ -\mathbf{I} & \mathbf{S}^U_{22} \end{pmatrix}^{-1} \begin{pmatrix} -\mathbf{S}^U_{11} & \mathbf{I} \\ -\mathbf{S}^U_{21} & 0 \end{pmatrix} \begin{pmatrix} \mathbf{A}^{(+)} \\ \mathbf{A}^{(-)} \end{pmatrix} = \lambda \begin{pmatrix} \mathbf{A}^{(+)} \\ \mathbf{A}^{(-)} \end{pmatrix} \quad (11)$$

It is convenient to normalize eigenvectors of Bloch states so that $|C_k| \equiv \left| \mathbf{A}_k^{(-)\dagger} \mathbf{A}_k^{(-)} - \mathbf{A}_k^{(+)\dagger} \mathbf{A}_k^{(+)} \right| = 1$. Those with $C_k = 1\ (-1)$ correspond to right- (left-) moving Bloch waves. Let us consider the eigenvectors $\mathbf{A}_j^{(\pm)}$ of right-moving Bloch waves $(j = 1, 2, \cdots, \tilde{Q})$ and left evanescence states $(j = \tilde{Q}+1, \cdots, Q)$. $\mathbf{S}^L$ relates the coefficients of incoming and outgoing waves in the coupled chains $(\mathbf{D}^{(\pm)})$ and in the lead $(\mathbf{L}^{(\pm)})$,

$$\begin{pmatrix} \mathbf{D}^{(-)} \\ \mathbf{L}^{(-)} \end{pmatrix} = \mathbf{S}^L \begin{pmatrix} \mathbf{D}^{(+)} \\ \mathbf{L}^{(+)} \end{pmatrix}. \quad (12)$$

$\mathbf{S}^L$ corresponds to the first $\tilde{Q}$ lines of the matrix [6]

$$\begin{pmatrix} \mathbf{M}_2^{(-)} \left[ \mathbf{M}_2^{(+)} \right]^{-1} & \mathbf{M}_1^{(-)} - \mathbf{M}_2^{(-)} \left[ \mathbf{M}_2^{(+)} \right]^{-1} \mathbf{M}_1^{(+)} \\ \left[ \mathbf{M}_2^{(+)} \right]^{-1} & -\left[ \mathbf{M}_2^{(+)} \right]^{-1} \mathbf{M}_1^{(+)} \end{pmatrix}, \quad (13)$$

with

$$\mathbf{M}_1^{(\pm)} = \begin{pmatrix} \mathbf{A}_1^{(\pm)} & \mathbf{A}_2^{(\pm)} & \cdots & \mathbf{A}_{\tilde{Q}}^{(\pm)} \end{pmatrix}, \quad (14)$$

and

$$\mathbf{M}_2^{(\pm)} = \begin{pmatrix} \mathbf{A}_1^{(\mp)*} & \cdots & \mathbf{A}_{\tilde{Q}}^{(\mp)*} & \mathbf{A}_{\tilde{Q}+1}^{(\pm)} & \cdots & \mathbf{A}_Q^{(\pm)} \end{pmatrix}. \quad (15)$$

In addition to inversion and multiplications of matrices, note that calculation of $\mathbf{S}^L$ requires to diagonalize matrix $\Lambda$ in Eq. (11). $\Lambda$ can also be expressed as a Taylor series of order $N$, and following the algorithm in Appendix B, its eigenvalues and eigenvectors can be computed as Taylor series of the same order, which leads us to the Taylor expansion of $\mathbf{S}^L$.

Lead structures can be glued to scattering regions using the RSMM [5]. In other words, S-matrices of arbitrary systems with general periodic leads can be expressed as Taylor series. Furthermore, the band structure can also be computed from the eigenvalues $(\lambda)$ as a Taylor series. Finally, since the transmission function is the trace of the product of submatrices of $\mathbf{S}$, Landauer conductance (1) can be computed as a Taylor series too.

## 3. Validation and convergence of Taylor expansions

Figure 2 shows the transmission function calculated with order $N = 30$ Taylor series (solid lines) for (a,b) atomic chains with two site $\varepsilon'$ impurities, (c,d) armchair and (e,f) zigzag nanoconstrictions (illustrated on top), considering different lengths of the scattering region $(L)$. $t$ is the hopping integral and site energy is zero in all cases, except for impurities whose site energy is $\varepsilon' = 0.5t$. Taylor series were expanded around energies denoted by open circles, and their relative errors are shown as dotted lines. Dark and bright colors facilitate the contrast between adjacent Taylor expansions. Transmission functions in instances 2(b), (d) and (e) agree to previously reported results. The behavior of relative errors confirms that Taylor series are valid in the neighborhood of each energy. However, variations in the number of channels lead to discontinuities in $T$ or in its derivatives, which are beyond a Taylor series approach. Moreover, relative errors are small only in segments that contain a fraction of oscillation of $T$. These segments are smaller if $L$ is bigger because a larger scattering region increases the number of these oscillations. Consequently, it is impossible to use a single Taylor series to approximate the full transmission function spectrum. Instead, several Taylor series should be calculated with different expansion centers.

Evaluation of a Taylor series at energy $E$ has an error $(R_N)$ given by the Lagrange form of the remainder,

$$R_N(E) = \frac{T^{(N+1)}(E^*)}{(N+1)!}(E-E_0)^{N+1} \qquad (16)$$

where $E_0$ is the expansion center of the series, $T^{(N+1)} = d^{N+1}T/dE^{N+1}$, and $E^*$ is an energy between $E_0$ and $E$. The order of magnitude of this error may be estimated by taking $T^{(N+1)}(E^*) \approx T^{(N+1)}(E_0)$, and then

$$|R_N(E)| \approx |T_{N+1}|\Delta E^{N+1} \qquad (17)$$

where $\Delta E = |E - E_0|$ and $T_{N+1}$ is the $(N+1)$-th coefficient in the series of the transmission function, which can be calculated with the RSMM.

In general, the range of transmission function is $[0, N_C]$, where $N_C$ is the number of open channels in the leads. On the other hand, truncated Taylor series are polynomials of degree $N$ that grow beyond the bounds $[0, N_C]$ as $(E - E_0)^N$ for large enough $\Delta E$. Consequently, if $|R_N|$ is of the order of $N_C$, Taylor series are not valid. Fig. 3(a) shows the value of $\Delta E$ as a function of $N$ that leads to an estimated error $|R_N| \approx N_C$ in Eq. (17), for the cases pointed by arrows in Fig. 2. Observe the asymptotic behavior as $N$ is increased, which is explained by a finite radius of convergence.

Figure 3(b) shows the magnitude of the transmission function evaluated in the complex domain by using the analytical expression reported by V. Sánchez *et al*. [7], with the same parameters of Fig. 2(b). Note the presence of complex-conjugated pole pairs. They cause the oscillations in $T(E)$. The maximum and width of each oscillation are related to the real and imaginary parts of each pole pair, respectively. This explains why the radius of convergence is limited to segments that contain only a fraction of oscillation in $T$. Dashed circles in Fig. 3(b) are plotted with radii equal to the asymptotic values of $\Delta E$ for each expansion center in Fig. 2(b). Their perimeter touches the poles, showing that the limiting behavior of $\Delta E$ gives a good estimation of the radius of convergence.

## 4. Conclusions

In this paper, we proposed a novel method to obtain arbitrary order Taylor series of scattering matrices and of Landauer conductance. It is based on the recursive scattering matrix method (RSMM), that allows us to calculate the scattering matrix of general multiterminal tight binding structures with arbitrary periodic leads. Calculation of the Taylor series is fully integrated within the recursive method, facilitating its implementation.

The method was validated in atomic chains with site defects and graphene nanoconstrictions. Calculated Taylor series in these systems give a good approximation of transmission functions for energies around the expansion center, being able to reach machine precision if the order of the Taylor series ($N$) is large enough. An estimation of the error caused by a truncated Taylor series was obtained by using the Lagrange form of the residue.

In general, Taylor series have finite convergence radius ($R$), due to poles in the complex domain. They produce the oscillations in the transmission function and reduce $R$ to fit in regions that contain only a fraction of an oscillation. Full transmission functions can be calculated from the Taylor series by taking a number of random expansion centers of the order of the number of oscillations.

Taylor series could improve the analysis of tight-binding structures, by increasing its computational efficiency, avoiding the calculation of the transmission function at too many energies. Moreover, they can be used to speed up the determination of Büttiker conductance in multiterminal devices, where the Newton-Raphson root-finder method could be employed to find the chemical potential such that the total current through each probe is zero [23].

Finally, we believe the Taylor series may be employed to determine the position of the poles, which could lead to a semi-analytical expression for the Landauer conductance. This study is currently under development.

**Acknowledgments**

This work was supported by UNAM-PAPIIT IN116819. Computations were performed at Miztli under project LANCAD-UNAM-DGTIC-329.

**Appendix A**

Let us consider two matrices, $\mathbf{M}^{(1)}$ and $\mathbf{M}^{(2)}$, given as Taylor expansions

$$\mathbf{M}^{(j)}(E) = \sum_{n=0}^{\infty} \mathbf{M}_n^{(j)} (E - E_0)^n \tag{A.1}$$

By straightforward calculation, the sum of these matrices is

$$\mathbf{\Sigma}(E) \equiv \mathbf{M}^{(1)} + \mathbf{M}^{(2)} = \sum_{n=0}^{\infty} \mathbf{\Sigma}_n (E - E_0)^n \qquad (A.2)$$

where

$$\mathbf{\Sigma}_n = \mathbf{M}_n^{(1)} + \mathbf{M}_n^{(2)}. \qquad (A.3)$$

On the other hand, the product of these matrices becomes

$$\mathbf{P}(E) \equiv \mathbf{M}^{(1)} \mathbf{M}^{(2)} = \sum_{n=0}^{\infty} \mathbf{P}_n (E - E_0)^n, \qquad (A.4)$$

with

$$\mathbf{P}_n = \sum_{m=0}^{n} \mathbf{M}_m^{(1)} \mathbf{M}_{n-m}^{(2)}. \qquad (A.5)$$

Moreover, the inverse of a matrix $\mathbf{A}$ becomes

$$\mathbf{B}(E) \equiv [\mathbf{A}(E)]^{-1} = \sum_{n=0}^{\infty} \mathbf{B}_n (E - E_0)^n \qquad (A.6)$$

where $\mathbf{B}_0 = \mathbf{A}_0^{-1}$ and

$$\mathbf{B}_n = -\sum_{m=0}^{n-1} \mathbf{A}_0^{-1} \mathbf{A}_{n-m} \mathbf{B}_m \qquad (A.7)$$

for $n \geq 1$. Notice from Eqs. (A.3), (A.5) and (A.7) that if Taylor series (A.1) is truncated to order N, then matrices $\mathbf{\Sigma}_n$, $\mathbf{P}_n$ and $\mathbf{B}_n$ can be exactly computed to the same order N.

**Appendix B**

Let us consider a $N \times N$ matrix $\mathbf{A}$ given as a Taylor series,

$$\mathbf{A}(E) = \sum_{n=0}^{\infty} \mathbf{A}_n (E - E_0)^n, \qquad (B.1)$$

where $\mathbf{A}_0$ is a diagonalizable matrix with no-degenerate eigenvalues $\lambda_m^{(0)}$ and corresponding left and right eigenvectors $\mathbf{u}_m^{(0)}$ and $\mathbf{v}_m^{(0)}$. Eigenvalues and right eigenvectors of $\mathbf{A}$ are given by

$$\lambda_m = \sum_{n=0}^{\infty} \lambda_m^{(n)} \left( E - E_0 \right)^n, \tag{B.2}$$

and

$$\mathbf{v}_m = \sum_{n=0}^{\infty} \mathbf{v}_m^{(n)} \left( E - E_0 \right)^n, \tag{B.3}$$

respectively. Following an analogous procedure to that of non-degenerate perturbation theory, coefficients of Taylor expansions (B.2) and (B.3) are given by

$$\lambda_m^{(n)} = \begin{cases} c_{m,m}^{(n)} & \text{if } n = 1 \\ c_{m,m}^{(n)} + \sum_{k=1}^{n-1} \sum_{p=1}^{N} b_{m,p}^{(k)} c_{p,m}^{(n-k)} & \text{if } n > 1 \end{cases}, \tag{B.4}$$

and

$$\mathbf{v}_m^{(n)} = \sum_{k=1}^{N} b_{m,k}^{(n)} \mathbf{v}_k^{(0)}, \tag{B.5}$$

where $b_{m,m}^{(n)} = 0$,

$$c_{n,m}^{(k)} = \frac{\mathbf{u}_m^{(0)} \mathbf{A}_k \mathbf{v}_n^{(0)}}{\mathbf{u}_m^{(0)} \mathbf{v}_m^{(0)}}, \tag{B.6}$$

and

$$b_{n,m}^{(k)} = \begin{cases} \dfrac{c_{n,m}^{(k)}}{\lambda_n^{(0)} - \lambda_m^{(0)}} & \text{if } k = 1 \\ \dfrac{c_{n,m}^{(k)} - \sum_{p=1}^{k-1} \left[ b_{n,m}^{(p)} \lambda_n^{(k-p)} - \sum_{j=1}^{N} b_{n,j}^{(p)} c_{j,m}^{(k-p)} \right]}{\lambda_n^{(0)} - \lambda_m^{(0)}} & \text{if } k > 1 \end{cases}. \tag{B.7}$$

It is important to mention that eigenvectors (B.3) are non-normalized, and their normalization coefficient must also be found as a Taylor series.

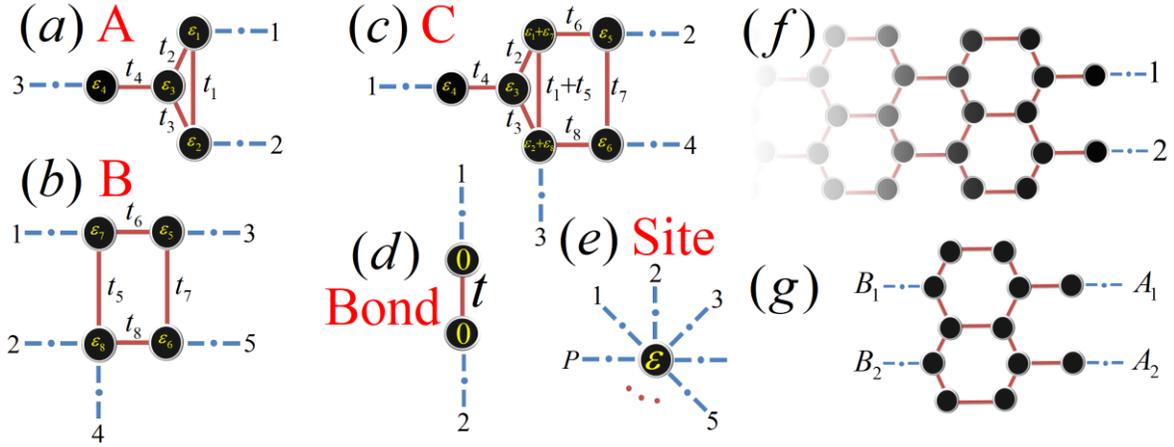

**Figure 1.** Tight-binding structures connected to semi-infinite atomic chains (blue dot dashed lines). Using the RSMM, system (c) is obtained by gluing systems (a) and (b) by making $A_n^{(\pm)} = B_n^{(\mp)}$ for $n = 1, 2$, where $n$ labels the coupled chains. RSMM allows us to model arbitrary structures starting from (d) bond and (e) site structures. (f) A lead structure obtained by using the unit structure (g).

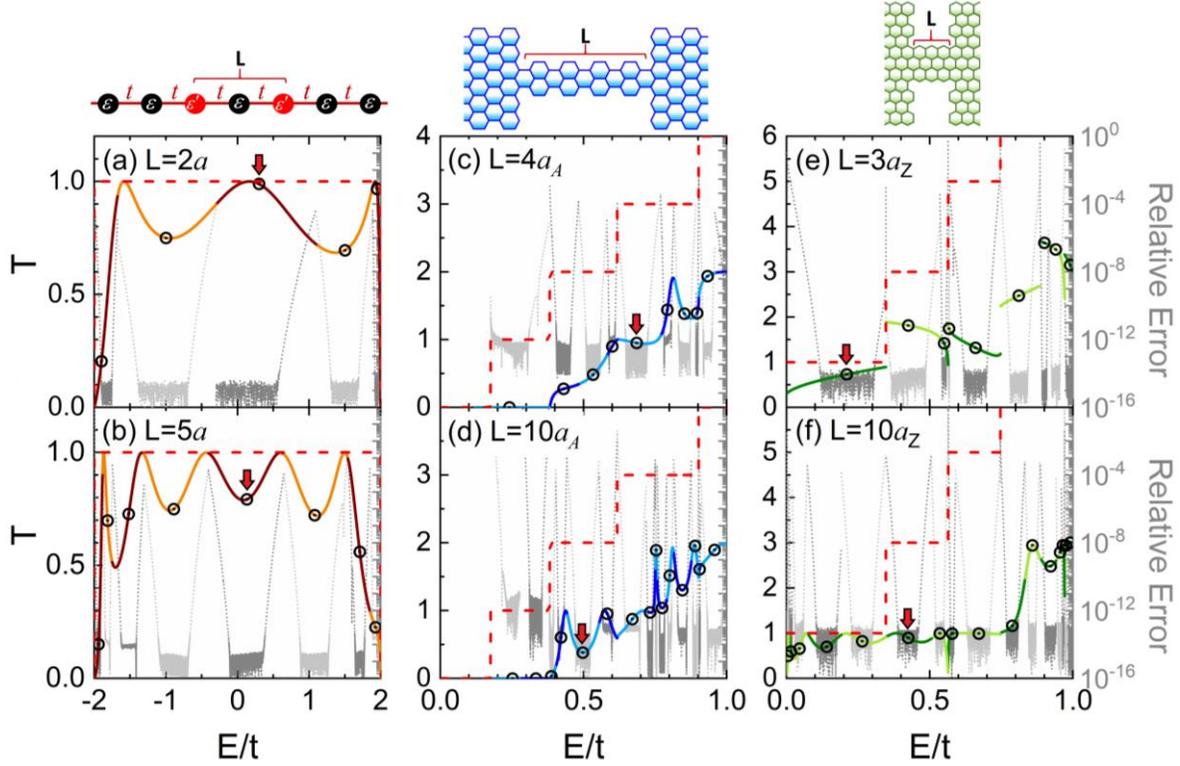

**Figure 2.** Transmission function (solid lines) and relative errors (dotted lines) obtained from order $N=30$ Taylor series within the RSMM, expanded around energies represented by open circles, for structures shown on top, (a,b) atomic chains with two site $\varepsilon' = 0.5t$ impurities, (c,d) armchair graphene nanoconstrictions and (e,f) zigzag graphene nanoconstrictions. $a$, $a_A$ and $a_Z$ are lattice parameters of the atomic chain, the armchair graphene nanoribbon and the zigzag graphene nanoribbon, respectively.

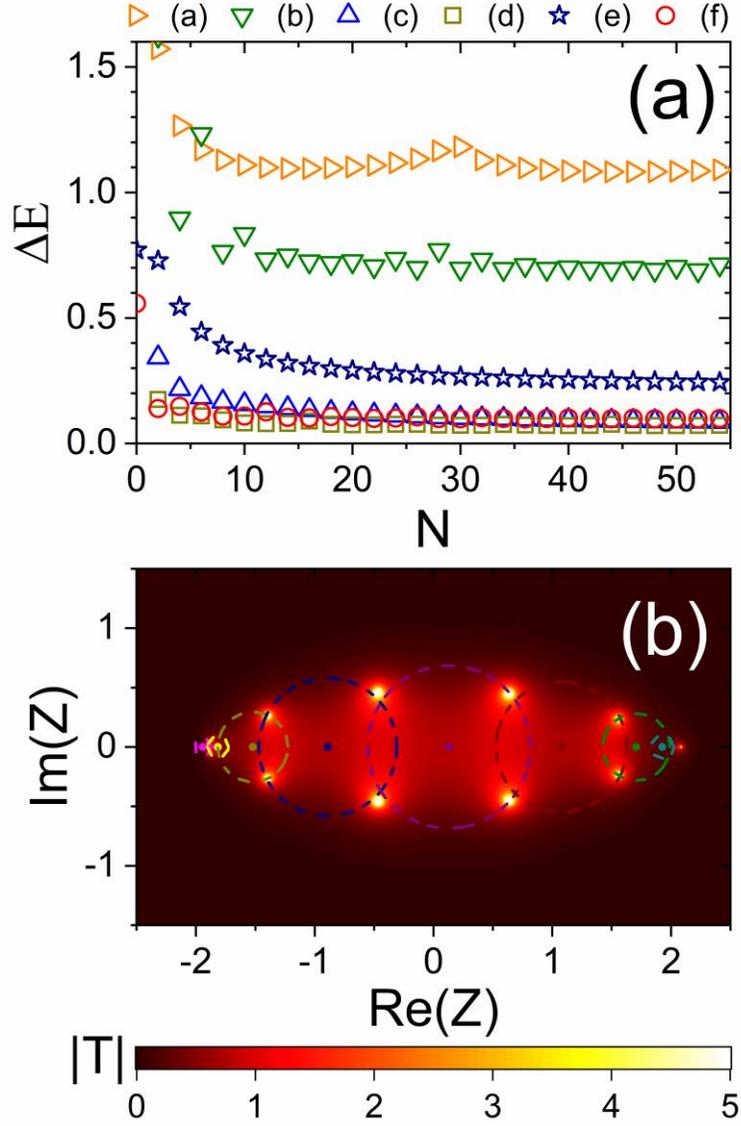

**Figure 3.** (a) $\Delta E$ as a function of $N$ that leads to an estimated error $|R_{N+1}| \approx N_C$ for the Taylor expansions pointed by arrows in Fig. 2. (b) Magnitude of the Transmission function, $|T|$, evaluated in the complex domain ($Z$), for the same parameters of Fig. 2(b); the radii of dashed circles are equal to asymptotic values of $\Delta E$ for each expansion center in Fig. 2(b).